\documentclass[aps,pra,twocolumn,nobibnotes,superscriptaddress,nofootinbib,10pt]{revtex4-1} 
\usepackage[latin1]{inputenc}
\usepackage{amsmath,amssymb}
\usepackage{mathrsfs} 
\usepackage{graphicx,color,colortbl}
\usepackage{rotating,array,tabularx,booktabs}
\usepackage{varwidth,xcolor}
\usepackage{placeins}
\usepackage{siunitx}
\usepackage[normalem]{ulem}

\DeclareGraphicsExtensions{.pdf,.PDF,.png,.PNG}
\allowdisplaybreaks

\newcommand{\dif}{\mathrm{d}}%
\newcommand{\ZT}[1]{\textquotedblleft#1\textquotedblright}%
\newcommand{\ww}{\vec{u}}%
\newcommand{\ws}{u}%

\begin{document}

\title{Acoustic propulsion of nano- and microcones: dependence on the viscosity of the surrounding fluid}

\author{Johannes Vo\ss{}}
\affiliation{Institut f\"ur Theoretische Physik, Center for Soft Nanoscience, Westf\"alische Wilhelms-Universit\"at M\"unster, 48149 M\"unster, Germany}

\author{Raphael Wittkowski}
\email[Corresponding author: ]{raphael.wittkowski@uni-muenster.de}
\affiliation{Institut f\"ur Theoretische Physik, Center for Soft Nanoscience, Westf\"alische Wilhelms-Universit\"at M\"unster, 48149 M\"unster, Germany}

\begin{abstract}
This article investigates how the acoustic propulsion of cone-shaped colloidal particles that are exposed to a traveling ultrasound wave depends on the viscosity of the fluid surrounding the particles. Using acoustofluidic computer simulations, we found that the propulsion of such nano- and microcones decreases strongly and even changes sign for increasing shear viscosity. In contrast, we found only a weak dependence of the propulsion on the bulk viscosity. The obtained results are in line with the findings of previous theoretical and experimental studies.
\begin{figure}[htb]
\centering
\fbox{\includegraphics[width=8cm]{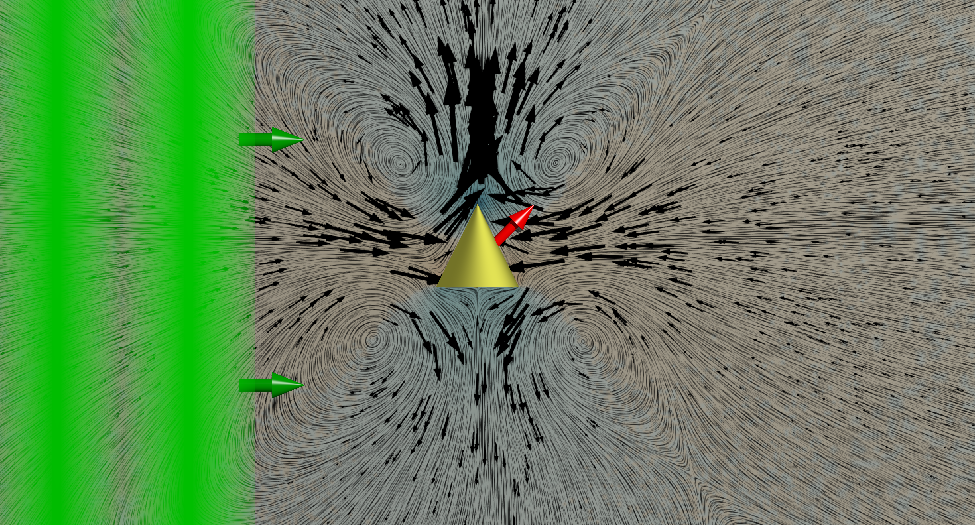}}%
\end{figure}
\end{abstract}
\maketitle

\section{Introduction}
After the idea of nano- and micromachines that carry out medical tasks inside a patient's body has been a dream for several decades \cite{Asimov1966}, the progress in nanotechnology at the end of the last century made the fabrication of motile nano- and microparticles (so-called active particles \cite{BechingerdLLRVV2016}) possible 
\cite{WangCHM2012,GarciaGradillaEtAl2013,AhmedEtAl2013,WangLMAHM2014,BalkEtAl2014,WuEtAl2014,GarciaGradillaSSKYWGW2014,AhmedGFM2014,Kiristi2015,WangDZSSM2015,AhmedLNLSMCH2015,EstebanFernandezdeAvilaMSLRCVMGZW2015,WuEtAl2015a,WuEtAl2015b,EstebanEtAl2016,SotoWGGGLKACW2016,KaynakONNLCH2016,AhmedWBGHM2016,AhmedBJPDN2016,UygunEtAl2017,KaynakONLCH2017,EstebanFernandezEtAl2017,RenZMXHM2017,HansenEtAl2018,EstebanEtAl2018,SabrinaTABdlCMB2018,WangGWSGXH2018,TangEtAl2019,ZhouZWW2017,GaoLWWXH2019,QualliotineEtAl2019,LuSZWPL2019,RenEtAl2019,ZhouYWDW2017,Zhou2018,DumyJMBGMHA2020,AghakhaniYWS2020,LiuR2020,ValdezLOESSWG2020,McneillSWOLNM2021,MohantyEtAl2021,LiMMOP2021,DaiEtAl2021,WangGSLGH2021,ChenEtAl2021,OuEtAl2021,ZhangFDSY2021,EbrahimiEtAl2021,ZhengEtAl2021,ZhouMMPZP2021}. 
During the last two decades, a large number of artificial motile nano- and microparticles that utilize various mechanisms for propulsion has been developed \cite{BechingerdLLRVV2016,Venugopalan2020,FernandezRMHSS2020,YangEtAl2020} and fascinating future applications of these particles have been envisaged in fields like medicine \cite{LiEFdAGZW2017,PengTW2017,SotoC2018,WangGZLH2020,WangZ2021}, where they could be used for targeted drug delivery \cite{LuoFWG2018,ErkocYCYAS2019,NitschkeW2021}, materials science \cite{Visser2007,JunH2010,McdermottKDKWSV2012,KuemmelSLVB2015,VanderMeerDF2016,JeanneretPKP2016,NeedlemanD2017,WangDZLGYLWM2019,RamananarivoDP2019,FratzlFKS2021}, where they could be used to form active crystals \cite{OphausGT2018,PraetoriusVWL2018,HollAGKOT2021,OphausKGT2020,OphausKGT2021,teVrugtJW2021} and other new types of matter \cite{EversW2022}, and environmental care \cite{SolerMFSS2013,GaoW2014a,SafdarKJ2018,GeCLLG2019,ChenEtAl2021,LiuZ2021,WangJWZ2021,ShivalkarEtAl2021}. 
 
Among all propulsion mechanisms that have been developed so far, acoustic propulsion \cite{WangCHM2012,GarciaGradillaEtAl2013,AhmedEtAl2013,WuEtAl2014,WangLMAHM2014,GarciaGradillaSSKYWGW2014,BalkEtAl2014,AhmedGFM2014,EstebanFernandezdeAvilaMSLRCVMGZW2015,WuEtAl2015a,WuEtAl2015b,RaoLMZCW2015,Kiristi2015,EstebanEtAl2016,KaynakONNLCH2016,SotoWGGGLKACW2016,AhmedWBGHM2016,AhmedBJPDN2016,UygunEtAl2017,KaynakONLCH2017,XuXZ2017,ZhouZWW2017,ZhouYWDW2017,EstebanFernandezEtAl2017,HansenEtAl2018,SabrinaTABdlCMB2018,WangGWSGXH2018,EstebanEtAl2018,LuSZWPL2019,QualliotineEtAl2019,GaoLWWXH2019,RenEtAl2019,VossW2020,ValdezLOESSWG2020,AghakhaniYWS2020,LiuR2020,DumyJMBGMHA2020,VossW2021,VossW2022orientation,VossW2022acoustic,FengYC2015,AhmedLNLSMCH2015,XieEtAl2015a,XieEtAl2015b,LaubliSAN2017,AhmedDHN2017,RenEtAl2019,AghakhaniYWS2020,McneillNBM2020,MohantyEtAl2021,McneillSWOLNM2021}, where nano- or microparticles in a fluid become motile when they are exposed to an ultrasound wave, belongs to the most promising mechanisms \cite{EstebanFernandezdeAvilaALGZW2018,SafdarKJ2018,PengTW2017,KaganBCCEEW2012,XuanSGWDH2018,XuGXZW2017,XuCLFPLK2019,RenWM2018,Zhou2018,FernandezRMHSS2020}. 
Important advantages of this mechanism compared to other ones are its biocompatibility \cite{WangLWXLGM2019,OuEtAl2020}, its compatibility with various types of fluids \cite{GarciaGradillaEtAl2013,WuEtAl2014,WangLMAHM2014,EstebanEtAl2018,GaoLWWXH2019,WangGZLH2020,WuEtAl2015a,EstebanFernandezdeAvilaMSLRCVMGZW2015,EstebanEtAl2016,EstebanFernandezEtAl2017,UygunEtAl2017,HansenEtAl2018,QualliotineEtAl2019}, and an easy way of permanently supplying the particles with energy \cite{XuGXZW2017,WangLWXLGM2019}. 
For special purposes, acoustic propulsion can even be combined with other propulsion mechanisms \cite{LILXKLWW2015,WangDZSSM2015,RenZMXHM2017,Zhou2018,RenWM2018,TangEtAl2019,ValdezLOESSWG2020}.  

Despite intensive research on acoustically propelled nano- and microparticles based on experiments \cite{WangCHM2012,GarciaGradillaEtAl2013,AhmedEtAl2013,WangLMAHM2014,BalkEtAl2014,WuEtAl2014,GarciaGradillaSSKYWGW2014,AhmedGFM2014,WangDZSSM2015,AhmedLNLSMCH2015,Kiristi2015,EstebanFernandezdeAvilaMSLRCVMGZW2015,WuEtAl2015a,WuEtAl2015b,EstebanEtAl2016,SotoWGGGLKACW2016,KaynakONNLCH2016,AhmedWBGHM2016,AhmedBJPDN2016,UygunEtAl2017,KaynakONLCH2017,EstebanFernandezEtAl2017,RenZMXHM2017,HansenEtAl2018,EstebanEtAl2018,SabrinaTABdlCMB2018,WangGWSGXH2018,TangEtAl2019,ZhouZWW2017,GaoLWWXH2019,QualliotineEtAl2019,RenEtAl2019,ZhouYWDW2017,Zhou2018,DumyJMBGMHA2020,AghakhaniYWS2020,LiuR2020,ValdezLOESSWG2020,McneillSWOLNM2021,MohantyEtAl2021}, computer simulations \cite{AhmedBJPDN2016,SabrinaTABdlCMB2018,Zhou2018,TangEtAl2019,VossW2020,VossW2021,VossW2022orientation,VossW2022acoustic}, and analytical approaches \cite{NadalL2014,CollisCS2017}, this is still a rapidly growing field of research with many important questions remaining unanswered.
For example, little is known about how the propulsion of such a particle depends on the viscosity of the surrounding fluid. 
This is, however, an important question, since in future applications the particles will be combined with various fluids \cite{GarciaGradillaEtAl2013,WuEtAl2014,WangLMAHM2014,EstebanEtAl2018,GaoLWWXH2019,WangGZLH2020,WuEtAl2015a,EstebanFernandezdeAvilaMSLRCVMGZW2015,EstebanEtAl2016,EstebanFernandezEtAl2017,UygunEtAl2017,HansenEtAl2018,QualliotineEtAl2019}.
 
Some experimental studies, that have compared the propulsion velocity of particles for a few surrounding fluids with different shear viscosities \cite{GarciaGradillaEtAl2013,WuEtAl2014,EstebanEtAl2018,GaoLWWXH2019,WangGZLH2020}, indicate that the propulsion velocity of the particles decreases for increasing shear viscosity.
A similar observation was made in another experiment where particles inside a cell moved slower than particles outside a cell \cite{WangLMAHM2014}. 
However, these experiments were not able to change the shear viscosity independently of other parameters (such as the bulk viscosity) of the fluid.
The experimental results are confirmed by a study that is based on an analytical approach \cite{CollisCS2017}, but the theory developed in this work completely neglects the bulk viscosity. 

In this article, we, therefore, address the viscosity dependence of the acoustic propulsion in more detail. 
We focus on cone-shaped nano- and microparticles that have been found to exhibit particularly strong propulsion \cite{VossW2021}. 
Our work is based on direct acoustofluidic simulations where we varied the fluid's shear viscosity as well as its bulk viscosity and studied how the values of these viscosities influence the flow fields generated around the particles and the particles' propulsion velocity.

\section{\label{methods}Methods}
For this study, we used the well-established methods described, e.g., in Ref.\ \cite{VossW2020}.

\subsection{Setup}
\begin{figure}[htb]
\centering
\includegraphics[width=\linewidth]{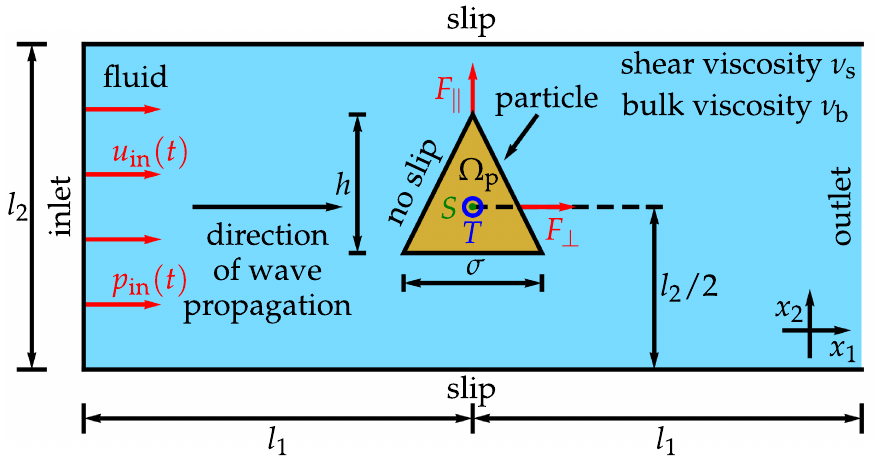}%
\caption{\label{fig:setup}Setup for the simulations.}
\end{figure}

The setup for our simulations is shown in Fig.\ \ref{fig:setup}.
It consists of a fluid-filled simulation domain with a rectangular shape that has width $2l_1$ (parallel to the $x_1$-axis) and height $l_2=\SI{200}{\micro\metre}$ (parallel to the $x_2$-axis).
The fluid is water that is initially at standard temperature $T_0=\SI{293.15}{\kelvin}$ and standard pressure $p_0=\SI{101325}{\pascal}$ and quiescent, i.e., it has a vanishing initial velocity field $\ww_0=\vec{0}\,\SI{}{\metre\,\second^{-1}}$. 
Its shear viscosity $\nu_\mathrm{s}\in [0.1,10]\,\SI{}{\milli\pascal\,\second}$ and bulk viscosity $\nu_\mathrm{b}\in [0.1,10]\,\SI{}{\milli\pascal\,\second}$ are varied.
A cone-shaped particle with diameter $\sigma=2^{-1/2}\SI{}{\micro\metre}$ and height $h=\sigma$ is positioned in the middle of the simulation domain such that the center of mass $\mathrm{S}$ of the particle domain $\Omega_\mathrm{p}$ coincides with the center of mass of the rectangular domain. 
Its orientation, which is given by the orientation of its axis of symmetry, is parallel to the $x_2$-axis. 
The shape of the particle is chosen analogously to Refs.\ \cite{VossW2021,VossW2022orientation,VossW2022acoustic}, since this shape has been found to lead to relatively efficient propulsion \cite{VossW2021} and this choice allows to compare directly with the previous work. 

At the left edge of the simulation domain, we prescribe inlet boundary conditions corresponding to a planar traveling ultrasound wave entering the system. 
For this purpose, we prescribe a time-dependent velocity $\ws_{\mathrm{in}}(t)=\Delta u \sin(2\pi f t)$ and pressure $p_{\mathrm{in}}(t)=\Delta p \sin(2\pi f t)$ at this edge. 
Here, $t$ is time, the velocity amplitude is given by $\Delta u=\Delta p / (\rho_0 c_{\mathrm{f}})$, and the pressure amplitude is given by $\Delta p=\SI{10}{\kilo\pascal}$.
We choose the frequency of the ultrasound wave as $f=\SI{1}{\MHz}$, the mass density of the unperturbed fluid as $\rho_0=\SI{998}{\kilogram\,\metre^{-3}}$, and the sound velocity as $c_\mathrm{f}=\SI{1484}{\metre\,\second^{-1}}$. 
To allow the wave to propagate in the $x_1$-direction without strong damping at the lower and upper edges of the simulation domain, we prescribe slip boundary conditions there. 
At the right edge of the simulation domain, we prescribe outlet boundary conditions so that the wave can leave the system.
For the boundary $\partial\Omega_\mathrm{p}$ of the particle, we choose no-slip boundary conditions.

The ultrasound wave has wavelength $\lambda=c_\mathrm{f}/f=\SI{1484}{\milli\metre}$ and acoustic energy density $E_\mathrm{R}=\Delta p^2/(2 \rho_0 c_{\mathrm{f}}^2)=\SI{22.7}{\milli\joule\,\metre^{-3}}$. 
Based on the wavelength $\lambda$, we choose the width of the simulation domain as $2l_1=\lambda/2$.
The interaction of the ultrasound wave with the particle leads to a propulsion force and a propulsion torque that act on the center of mass $\mathrm{S}$ of the particle. 
We denote the time-averaged stationary propulsion force as $\vec{F}$, which can be decomposed into a component $F_\parallel$ parallel to the particle and a component $F_\perp$ perpendicular to the particle, and the time-averaged stationary propulsion torque as $T$.

\subsection{Parameters}
All parameters that are relevant for the simulations and the values chosen for them are summarized in Table \ref{tab:Parameters}. For consistency, the values are chosen analogously to Ref.\ \cite{VossW2020}.
\begin{table*}[htb]
\centering
\caption{\label{tab:Parameters}Simulation parameters and their values. For consistency, the values are analogous to Ref.\ \cite{VossW2020}. The values of $c_\mathrm{f}$ and $\rho_0$ correspond to quiescent water at temperature $T_0$ and pressure $p_0$.}%
\begin{ruledtabular}
\begin{tabular}{@{}l@{}c@{}c@{}}%
\textbf{Name} & \textbf{Symbol} & \textbf{Value}\\
\hline
Particle diameter & $\sigma$ & $2^{-1/2}\SI{}{\micro\metre}$\\
Particle height & $h$ & $\sigma$\\
Sound frequency & $f$ & $\SI{1}{\MHz}$\\
Speed of sound & $c_\mathrm{f}$ & $\SI{1484}{\metre\,\second^{-1}}$\\
Time period of sound & $\tau=1/f$ & $\SI{1}{\micro\second}$\\
Wavelength of sound & $\lambda=c_\mathrm{f}/f$ & $\SI{1.484}{\milli\metre}$\\
Temperature of fluid & $T_0$ & $\SI{293.15}{\kelvin}$ (normal temperature)\\
Mean mass density of fluid & $\rho_0$ & $\SI{998}{\kilogram\,\metre^{-3}}$\\
Mean pressure of fluid & $p_{0}$ & $\SI{101325}{\pascal}$ (normal pressure) \\
Initial velocity of fluid & $\ww_{0}$ & $\vec{0}\,\SI{}{\metre\,\second^{-1}}$ \\
Sound pressure amplitude & $\Delta p$ & $\SI{10}{\kilo\pascal}$\\
Flow velocity amplitude & $\Delta u=\Delta p / (\rho_0 c_\mathrm{f})$ & $\SI{6.75}{\milli\metre\,\second^{-1}}$ \\
Acoustic energy density & $E_\mathrm{R}=\Delta p^2/(2 \rho_0 c_{\mathrm{f}}^2)$ & $\SI{22.7}{\milli\joule\,m^{-3}}$\\
Shear/dynamic viscosity of fluid & $\nu_{\mathrm{s}}$ & $\SI{0.1}{}$-$\SI{10}{\milli\pascal\,\second}$ \\
Bulk/volume viscosity of fluid & $\nu_{\mathrm{b}}$ & $\SI{0.1}{}$-$\SI{10}{\milli\pascal\,\second}$ \\
Inlet-particle distance & $l_{1}$ & $\lambda/4$ \\
Domain width & $l_2$ & $\SI{200}{\micro\metre}$\\
Mesh-cell size & $\Delta x$ & $\SI{15}{\nano \metre}$-$\SI{1}{\micro \metre}$ \\
Time-step size & $\Delta t$ & $1$-$\SI{10}{\pico \second}$\\
Simulation duration & $t_{\mathrm{max}}$ & $\geqslant 500\tau$ \\
Euler number & $\mathrm{Eu}$ & $\SI{2.20}{\cdot10^5}$ \\
Helmholtz number & $\mathrm{He}$ & $\SI{4.76}{\cdot10^{-4}}$ \\
Bulk Reynolds number &  $\mathrm{Re}_\mathrm{b}$ & $\SI{4.76}{\cdot10^{-4}}$-$\SI{4.76}{\cdot10^{-2}}$\\
Shear Reynolds number &  $\mathrm{Re}_\mathrm{s}$ & $\SI{4.76}{\cdot10^{-4}}$-$\SI{4.76}{\cdot10^{-2}}$\\
Particle Reynolds number &  $\mathrm{Re}_\mathrm{p}$ & $<\SI{}{10^{-4}}$ 
\end{tabular}%
\end{ruledtabular}%
\end{table*}

\subsection{Acoustofluidic simulations}
In our direct computational fluid dynamics simulations, we solve the continuity equation for the fluid's mass-density field, the compressible Navier-Stokes equations, and a linear constitutive equation for the pressure field with the finite volume software package OpenFOAM \cite{WellerTJF1998}. 
Nondimensionalization of the equations leads to four dimensionless numbers. These are the Euler number Eu, the Helmholtz number $\mathrm{He}$, a bulk Reynolds number $\mathrm{Re}_\mathrm{b}$, and a shear Reynolds number $\mathrm{Re}_\mathrm{s}$:
\begin{align}
\mathrm{Eu}&=\frac{\Delta p}{\rho_0 \Delta u^2}\approx \SI{2.20}{\cdot10^5},\\
\mathrm{He}&=\frac{f \sigma}{c_\mathrm{f}}\approx \SI{4.76}{\cdot10^{-4}},\\
\mathrm{Re}_\mathrm{b}&=\frac{\rho_0 \Delta u \sigma}{\nu_\mathrm{b}}\approx \SI{4.76}{\cdot10^{-4}}\mathrm{-}\SI{4.76}{\cdot10^{-2}},\\
\mathrm{Re}_\mathrm{s}&=\frac{\rho_0 \Delta u \sigma}{\nu_\mathrm{s}}\approx \SI{4.76}{\cdot10^{-4}}\mathrm{-}\SI{4.76}{\cdot10^{-2}}.
\end{align}
A detailed discussion of the meaning of these dimensionless numbers can be found in Ref.\ \cite{VossW2021}. 

When solving the field equations with the finite volume method, we use a structured, mixed rectangular-triangular mesh.
It has about 300,000 cells with a position-dependent cell size $\Delta x$ that is very small near the particle and larger further away from it.
To increase the performance of the simulations, we use an adaptive time-step method with a variable time-step size $\Delta t$. We ensure that $\Delta t$ always fulfills the Courant-Friedrichs-Lewy condition 
\begin{align}
C = c_\mathrm{f} \frac{\Delta t}{\Delta x} < 1 .
\end{align}
We simulate the system for a duration of $t_\mathrm{max} \geqslant 500\tau$ with the period $\tau$ of the ultrasound wave.
Due to the necessary fine discretization in space and time, the computational expense for an individual simulation was about $36,000$ CPU core hours.

\subsection{Propulsion force and torque}
To calculate the time-averaged stationary propulsion force $\vec{F}$, with components $F_\parallel$ and $F_\perp$, and the 
time-averaged stationary propulsion torque $T$ in the laboratory frame, we first determine the time-dependent force and torque that are exerted on the particle. 
For this purpose, we calculate from the space- and time-dependent velocity and pressure fields of the fluid, which are obtained from the acoustofluidic simulations, the quantities \cite{LandauL1987}
{\allowdisplaybreaks\begin{align}%
F^{(\alpha)}_{i} &=  \sum^{2}_{j=1} \int_{\partial\Omega_{\mathrm{p}}} \!\!\!\!\!\!\! \Sigma^{(\alpha)}_{ij}\,\dif A_{j}, 
\label{eq:F}\\%
T^{(\alpha)} &=  \sum^{2}_{j,k,l=1} \int_{\partial\Omega_{\mathrm{p}}} \!\!\!\!\!\!\! \epsilon_{3jk}(x_j-x_{\mathrm{p},j})\Sigma^{(\alpha)}_{kl}\,\dif A_{l}  
\label{eq:T}%
\end{align}}%
with $\alpha\in\{p,v\}$. 
These are the pressure component (superscript \ZT{$(p)$}) and viscous component (superscript \ZT{$(v)$}) of the time-dependent force $\vec{F}^{(p)}+\vec{F}^{(v)}$ and torque $T^{(p)}+T^{(v)}$ that act on the particle. 
Here, $\Sigma^{(p)}$ and $\Sigma^{(v)}$ are the pressure component and the viscous component of the stress tensor $\Sigma$, respectively, $\dif\vec{A}(\vec{x})=(\dif A_{1}(\vec{x}),\dif A_{2}(\vec{x}))^{\mathrm{T}}$ is the normal (outwards oriented) surface element of $\partial\Omega_{\mathrm{p}}$ at position $\vec{x}\in\partial\Omega_{\mathrm{p}}$, $\epsilon_{ijk}$ is the Levi-Civita symbol, and $\vec{x}_{\mathrm{p}}$ is the position of the center of mass $\mathrm{S}$.

The time-averaged stationary propulsion force $\vec{F}$ and torque $T$ are then obtained by locally averaging the time-dependent quantities over one period and extrapolating towards $t \to \infty$ with the method presented in Ref.\ \cite{VossW2020}.
This yields the propulsion force $\vec{F}=\vec{F}_p+\vec{F}_v$ with pressure component $\vec{F}_p=\langle\vec{F}^{(p)}\rangle$, where $\langle\cdot\rangle$ is the time average, and viscous component $\vec{F}_v=\langle\vec{F}^{(v)}\rangle$ and the propulsion torque $T=T_p + T_v$ with pressure component $T_p=\langle T^{(p)}\rangle$ and viscous component $T_v=\langle T^{(v)}\rangle$.
The components $F_\parallel$ and $F_\perp$ are then obtained by considering the Cartesian elements of $\vec{F}$ individually. 
We obtain the parallel force $F_\parallel=(\vec{F})_2=F_{\parallel,p}+F_{\parallel,v}$ with pressure component $F_{\parallel,p}=(\langle\vec{F}^{(p)}\rangle)_2$ and viscous component $F_{\parallel,v}=(\langle\vec{F}^{(v)}\rangle)_2$ as well as the perpendicular force $F_\perp = (\langle\vec{F}_\perp\rangle)_1=F_{\perp,p}+F_{\perp,v}$ with pressure component $F_{\perp,p}=(\langle\vec{F}^{(p)}\rangle)_1$ and viscous component $F_{\perp,v}=(\langle\vec{F}^{(v)}\rangle)_1$.

\subsection{Translational and angular propulsion velocity}
Introducing the force-torque vector $\vec{\mathfrak{F}}=(\vec{F},T)^{\mathrm{T}}$ and the translational-angular velocity vector $\vec{\mathfrak{v}}=(\vec{v},\omega)^{\mathrm{T}}$, we can transform from the time-averaged stationary propulsion force $\vec{F}$ and torque $T$ that act on the particle to the corresponding translational propulsion velocity $\vec{v}$, with components $v_{\parallel}=(\vec{v})_2$ and $v_{\perp}=(\vec{v})_1$, and angular propulsion velocity $\omega$ via the Stokes law \cite{HappelB1991}
\begin{equation}
\vec{\mathfrak{v}}=\frac{1}{\nu_\mathrm{s}}\boldsymbol{\mathrm{H}}^{-1}\,\vec{\mathfrak{F}}. 
\label{eq:velocity}%
\end{equation}
Here, the hydrodynamic resistance matrix
\begin{equation}
\boldsymbol{\mathrm{H}}=
\begin{pmatrix}
\boldsymbol{\mathrm{K}} & \boldsymbol{\mathrm{C}}^{\mathrm{T}}_{\mathrm{S}} \\
\boldsymbol{\mathrm{C}}_{\mathrm{S}} & \boldsymbol{\Omega}_{\mathrm{S}} 
\end{pmatrix} 
\label{eq:H}%
\end{equation}
with submatrices $\boldsymbol{\mathrm{K}}$, $\boldsymbol{\mathrm{C}}_{\mathrm{S}}$, and $\boldsymbol{\Omega}_{\mathrm{S}}$ depends on the size and shape of the particle.  
The subscript $\mathrm{S}$ at $\boldsymbol{\mathrm{C}}_{\mathrm{S}}$ and $\boldsymbol{\Omega}_{\mathrm{S}}$ denotes a reference point on which these submatrices depend. This reference point is chosen here as the center of mass $\mathrm{S}$.

Since $\boldsymbol{\mathrm{H}}$ needs to be calculated for a three-dimensional particle, whereas our acoustofluidic simulations are performed in two spatial dimensions (to reduce the computational effort to a manageable amount), we ascribe a thickness of $\sigma$ in the third dimension to the particle and calculate $\boldsymbol{\mathrm{H}}$ for the resulting three-dimensional particle.
Using the software \texttt{HydResMat} \cite{VossW2018,VossJW2019}, we then obtained the submatrices
\begin{align}
\boldsymbol{\mathrm{K}} &= \begin{pmatrix}
\SI{7.74}{\micro\metre} & 0 & 0\\
0 & \SI{7.48}{\micro\metre} & 0 \\
0 & 0 & \SI{7.16}{\micro\metre}
\end{pmatrix},
\label{eq:K}
\\
\boldsymbol{\mathrm{C}}_{\mathrm{S}} &= \begin{pmatrix}
0 & 0 & \SI{0.05}{\micro\metre^2}\\
0 & 0 & 0 \\
\SI{-0.11}{\micro\metre^2} & 0 & 0
\end{pmatrix},
\\
\boldsymbol{\Omega}_{\mathrm{S}} &= \begin{pmatrix}
\SI{1.81}{\micro\metre^3} & 0 & 0\\
0 & \SI{1.69}{\micro\metre^3} & 0 \\
0 & 0 & \SI{1.73}{\micro\metre^3}
\end{pmatrix}.
\label{eq:Omega}
\end{align}
To be able to work with these 3$\times$3-dimensional matrices, we use the three-dimensional versions of Eqs.\ \eqref{eq:F}-\eqref{eq:velocity} and neglect the contributions $\mathrm{K_{33}}$, $\mathrm{C_{13}}$, $\mathrm{\Omega_{11}}$, and $\mathrm{\Omega_{22}}$, which correspond to the particle's lower and upper surface.

\subsection{\label{sec:characterization}Characterization of particle motion}
The particle Reynolds number
\begin{align}
\mathrm{Re}_\mathrm{p}=\frac{\rho_0\sigma}{\nu_\mathrm{s}}\sqrt{v_\parallel^2+v_\perp^2} 
<10^{-4}
\end{align}
shows that the particle's motion through the fluid is dominated by viscous forces. 
To further characterize the motion of the particle, we compare its Brownian rotation with its translational and rotational propulsion \cite{VossW2021}. 
Depending on the acoustic energy density $E$ (in our simulations, we use $E=E_\mathrm{R}$), to which the particle's propulsion is approximately proportional \cite{VossW2022acoustic}, we can distinguish the following types of motion \cite{VossW2021}: 
\begin{itemize}
\item ${E < \min\{E_\mathrm{dir},E_\mathrm{gui}\}}$:\quad Random motion, 
\item ${E > \min\{E_\mathrm{dir},E_\mathrm{gui}\}}$:\quad Directional motion,
\item ${E < E_\mathrm{gui}}$:\qquad\qquad\qquad Random orientation,
\item ${E > E_\mathrm{gui}}$:\qquad\qquad\qquad Guided motion.
\end{itemize}
\ZT{Random motion} means that Brownian rotation dominates translational and rotational propulsion.
\ZT{Directional motion}, on the other hand, means that translational or rotational (which can align the orientation of the particle \cite{VossW2022orientation}) propulsion dominates Brownian rotation.
Furthermore, \ZT{Random orientation} means that Brownian rotation dominates rotational propulsion. 
\ZT{Guided motion}, on the other hand, means that rotational propulsion dominates Brownian rotation. 
The energy density thresholds $E_\mathrm{dir}$ and $E_\mathrm{gui}$ are defined as \cite{VossW2021}
\begin{align}
E_\mathrm{dir} &= \frac{\sigma D_\mathrm{R} E_\mathrm{R}}{|v_\parallel|}, \label{eq:E_dir_lim} \\
E_\mathrm{gui} &= \frac{\pi D_\mathrm{R} E_\mathrm{R}}{2|\omega|} \label{eq:E_Br_lim}
\end{align}
with the particle's rotational diffusion coefficient $D_\mathrm{R}=(k_\mathrm{B} T_0 / \nu_\mathrm{s}) (\boldsymbol{\mathrm{H}}^{-1})_{66}$ that corresponds to a rotation in the $x_1$-$x_2$ plane.
Here, $k_\mathrm{B}$ denotes the Boltzmann constant.

\section{\label{results}Results and discussion}
To study how the acoustic propulsion of a cone-shaped particle in a planar traveling ultrasound wave (see Fig.\ \ref{fig:setup}) depends on the viscosity of the surrounding fluid, we calculated the propulsion of such a particle for various values of the shear viscosity $\nu_\mathrm{s}$ and the bulk viscosity $\nu_\mathrm{b}$ of the fluid. 
We consider a variation of $\nu_\mathrm{s}$ while $\nu_\mathrm{b}$ is kept constant, a variation of $\nu_\mathrm{b}$ while $\nu_\mathrm{s}$ is kept constant, and a joint variation of $\nu_\mathrm{s}$ and $\nu_\mathrm{b}$. 
The viscosities are varied in the interval $[0.1,10]\,\SI{}{\milli\pascal\,\second}$. 
As reference viscosities, whose values are chosen when a viscosity is kept constant, we use those of water at standard temperature $T_0=\SI{293.15}{\kelvin}$ and standard pressure $p_0=\SI{101325}{\pascal}$, i.e., $\nu_\mathrm{s}=\SI{1}{\milli\pascal\,\second}$ and $\nu_\mathrm{b}=\SI{2.87}{\milli\pascal\,\second}$. 

We study the effect of the viscosities on the time-averaged stationary flow fields generated around the particle as well as on the strength and direction of the particle's propulsion. 
As relevant quantities for a characterization of the strength and direction of the propulsion, we consider the time-averaged stationary propulsion force $F_\parallel$ parallel to the particle's orientation (as defined in Fig.\ \ref{fig:setup}), the time-averaged stationary propulsion force $F_\perp$ perpendicular to the particle's orientation, the time-averaged stationary propulsion torque $T$ that tends to rotate the particle within the $x_1$-$x_2$ plane, their pressure components $F_{\parallel,p}$, $F_{\perp,p}$, and $T_p$, their viscous components $F_{\parallel,v}$, $F_{\perp,v}$, and $T_v$, as well as the translational propulsion velocities $v_\parallel$ and $v_\perp$, which correspond to $F_\parallel$ and $F_\perp$, respectively, and the angular propulsion velocity $\omega$ that corresponds to $T$.
After the analysis of these relevant quantities, we compare the propulsion torque of the particle with its rotational diffusion coefficient and study on this basis which type of motion the particle can be expected to exhibit depending on the viscosities and the acoustic energy density of the ultrasound.

\subsection{\label{sec:flowfields}Viscosity-dependent flow fields}
\begin{figure*}[htb]
\centering
\includegraphics[width=\linewidth]{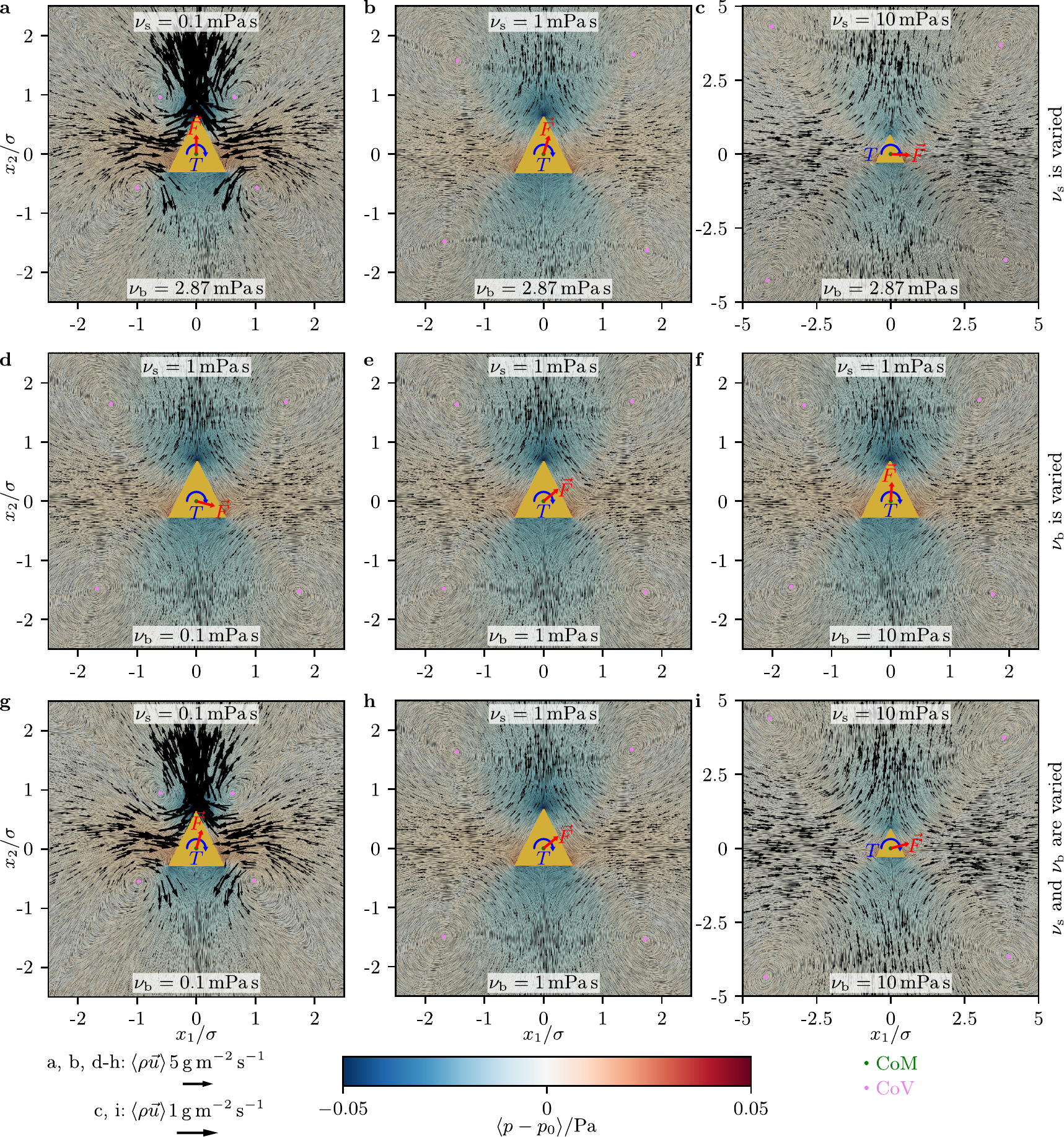}%
\caption{\label{fig:flowfield}Time-averaged mass-current density $\langle\rho\ww\rangle$ and reduced pressure $\langle p-p_{0}\rangle$ for varying \textbf{a}-\textbf{c} shear viscosity $\nu_\mathrm{s}$, \textbf{d}-\textbf{f} bulk viscosity $\nu_\mathrm{b}$, and \textbf{g}-\textbf{i} shear and bulk viscosities $\nu_\mathrm{s}$ and $\nu_\mathrm{b}$.
The center of mass (CoM) of the particle, the centers of vortices (CoV) of the flow field, and the directions of the particle's propulsion force $\vec{F}$ and propulsion torque $T$ are indicated.}
\end{figure*}

Our simulation results for the time-averaged stationary flow fields generated around the particle are shown in Fig.\ \ref{fig:flowfield}.
The qualitative structure of the flow field is the same for all considered values of the viscosities $\nu_\mathrm{s}$ and $\nu_\mathrm{b}$ and it is consistent with the flow fields of acoustically propelled particles that have been presented previously in the literature \cite{VossW2020,VossW2021,VossW2022orientation,VossW2022acoustic}. 
It includes 4 vortices that are placed at the top left, top right, bottom left, and bottom right of the particle. 
The vortices cause the fluid to flow on the left and right towards the particle and on the bottom and top away from it. 
Accordingly, the pressure of the fluid is increased on the left and right of the particle and it is reduced on the bottom and top of the particle. 

\begin{figure*}[htb]
\centering
\includegraphics[width=\linewidth]{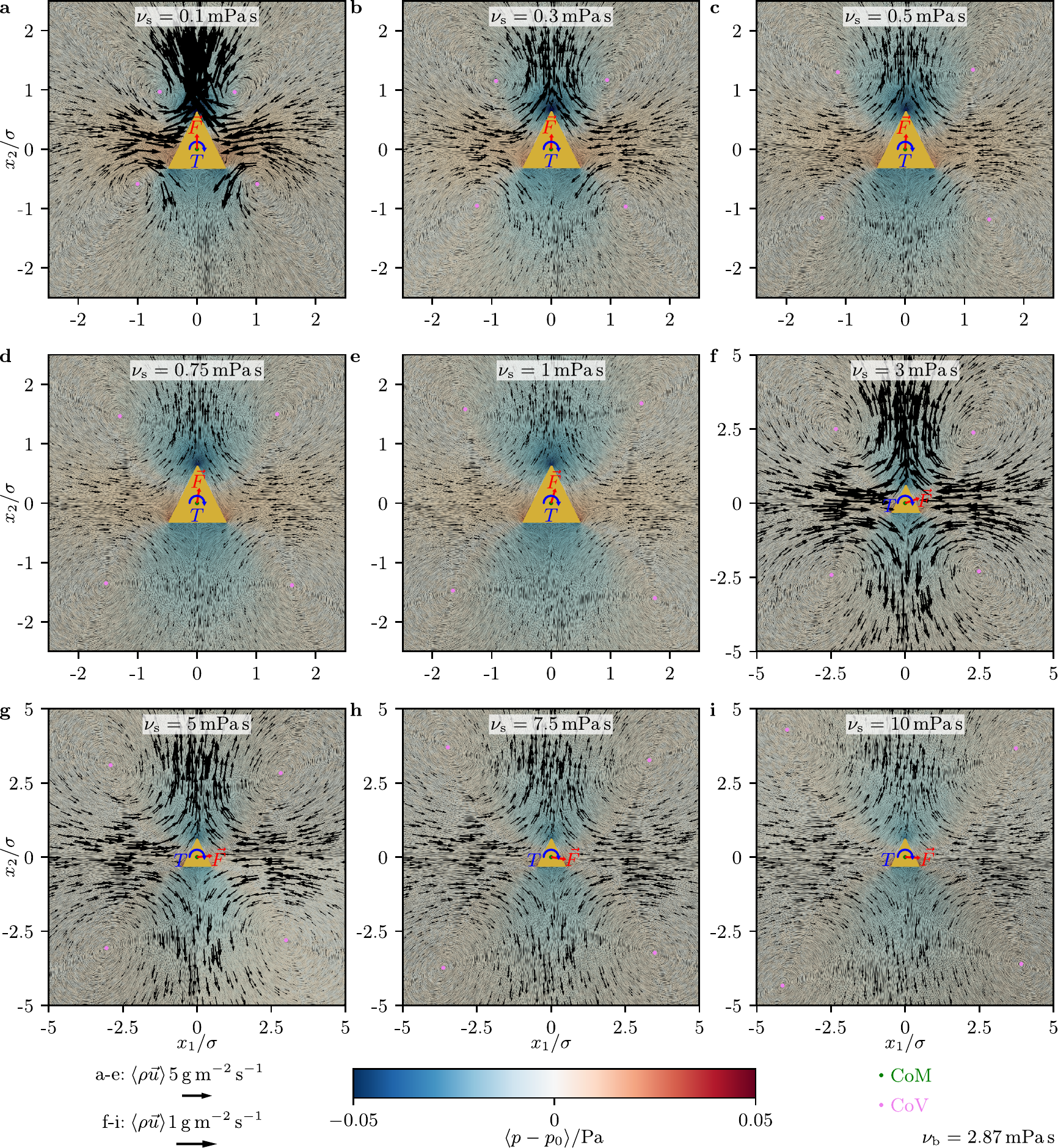}%
\caption{\label{fig:flowfield_s}The same as in Fig.\ \ref{fig:flowfield}\textbf{a}-\textbf{c}, but now for more values of the shear viscosity $\nu_\mathrm{s}$.}
\end{figure*}

When the shear viscosity $\nu_\mathrm{s}$ is increased (see Fig.\ \ref{fig:flowfield}\textbf{a}-\textbf{c} and Fig.\ \ref{fig:flowfield_s}), the vortices move away from the particle. 
As a consequence, the flow near the particle becomes weaker. The size of the regions with reduced or increased pressure, on the other hand, remains roughly the same. 
When we increase the bulk viscosity $\nu_\mathrm{b}$ (see Fig.\ \ref{fig:flowfield}\textbf{d}-\textbf{f}), we see only a slight shift of the vortices away from the particle and no significant change in the flow field. 
Furthermore, changing both viscosities $\nu_\mathrm{s}$ and $\nu_\mathrm{b}$ jointly (see Fig.\ \ref{fig:flowfield}\textbf{g}-\textbf{i}) leads to similar flow fields as for changing only the shear viscosity $\nu_\mathrm{s}$.
From these observations, we can expect that the strength of the particle's propulsion will decrease for increasing $\nu_\mathrm{s}$ and that it will only weakly depend on $\nu_\mathrm{b}$. 

\begin{figure}[tb]
\centering
\includegraphics[width=\linewidth]{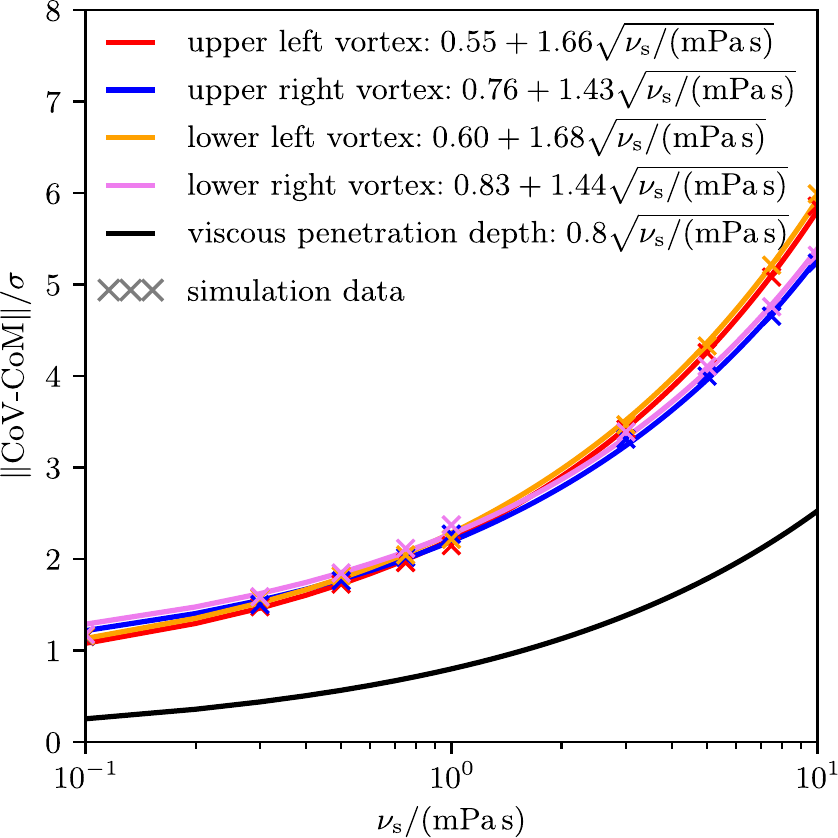}%
\caption{\label{fig:CoV}Distance of the centers of the vortices (CoV) from the particle's center of mass (CoM) for the flow field shown in Fig.\ \ref{fig:flowfield_s}.}
\end{figure}

Since the positions of the vortices have a strong influence on the structure of the flow field, we show in Fig.\ \ref{fig:CoV} the distance of the vortices from the center of mass of the particle as a function of the shear viscosity $\nu_\mathrm{s}$.
One can see that all vortices have a similar distance from the particle for low values of $\nu_\mathrm{s}$ and that the vortices move away from the particle when $\nu_\mathrm{s}$ increases, where the left vortices move slightly faster than the right vortices.
Since for low values of $\nu_\mathrm{s}$, the left vortices are a bit closer to the particle than the right vortices, there is a viscosity $\nu_\mathrm{s}\approx \SI{1}{\milli\pascal\,\second}$ where the distance from the particle is the same for the left and right pairs of vortices. 
The viscosity dependence of the vortex-to-particle distance can be described by a function of the form 
\begin{equation}
\frac{\mathfrak{f}(\nu_\mathrm{s})}{\sigma} = a + b \sqrt{\nu_\mathrm{s}/(\SI{}{\milli\pascal\,\second})}
\label{eq:fitvortices}%
\end{equation}
with the particle's diameter $\sigma$ and coefficients $a$ and $b$.
Fitting this function to the simulation data for the vortex-to-particle distance results in the fit functions that are given and visualized in Fig.\ \ref{fig:CoV}. 
A comparison with the simulation data shows that the agreement is excellent. 

The form of the function \eqref{eq:fitvortices} can be related to the viscosity dependence of boundary-layer-driven acoustic streaming.  
When ultrasound interacts with a boundary, Schlichting vortex streaming within a viscous boundary layer can emerge \cite{Boluriaan2003}. 
The typical thickness of this viscous boundary layer is given by $2\delta(\nu_\mathrm{s})$ with the viscous penetration depth \cite{LandauL1987,WiklundRO2012} 
\begin{equation}
\delta(\nu_\mathrm{s})=\sqrt{\frac{\nu_\mathrm{s}}{\pi\rho_0 f}} 
\approx 0.8 \sqrt{\nu_\mathrm{s}/(\SI{}{\milli\pascal\,\second})} \sigma .
\end{equation}
Interestingly, the scaling of the viscous penetration depth $\delta$ with the shear viscosity $\nu_\mathrm{s}$ is the same as in Eq.\ \eqref{eq:fitvortices} for the vortex-to-particle distance.
Averaging the functions for the vortex-to-particle distance over the 4 vortices and neglecting the offset, we find that the mean vortex-to-particle distance scales as $\approx 1.55 \sqrt{\nu_\mathrm{s}/(\SI{}{\milli\pascal\,\second})}\sigma \approx 2\delta$. 
This indicates that the acoustic propulsion mechanism is strongly linked to acoustic streaming. 
In previous work, a similar scaling as found here for the viscosity dependence of the vortex-to-particle distance was found for its dependence on the ultrasound frequency $f$ \cite{VossW2022acoustic}.

\subsection{Viscosity-dependent propulsion}
\begin{figure*}[htb]
\centering
\includegraphics[width=\linewidth]{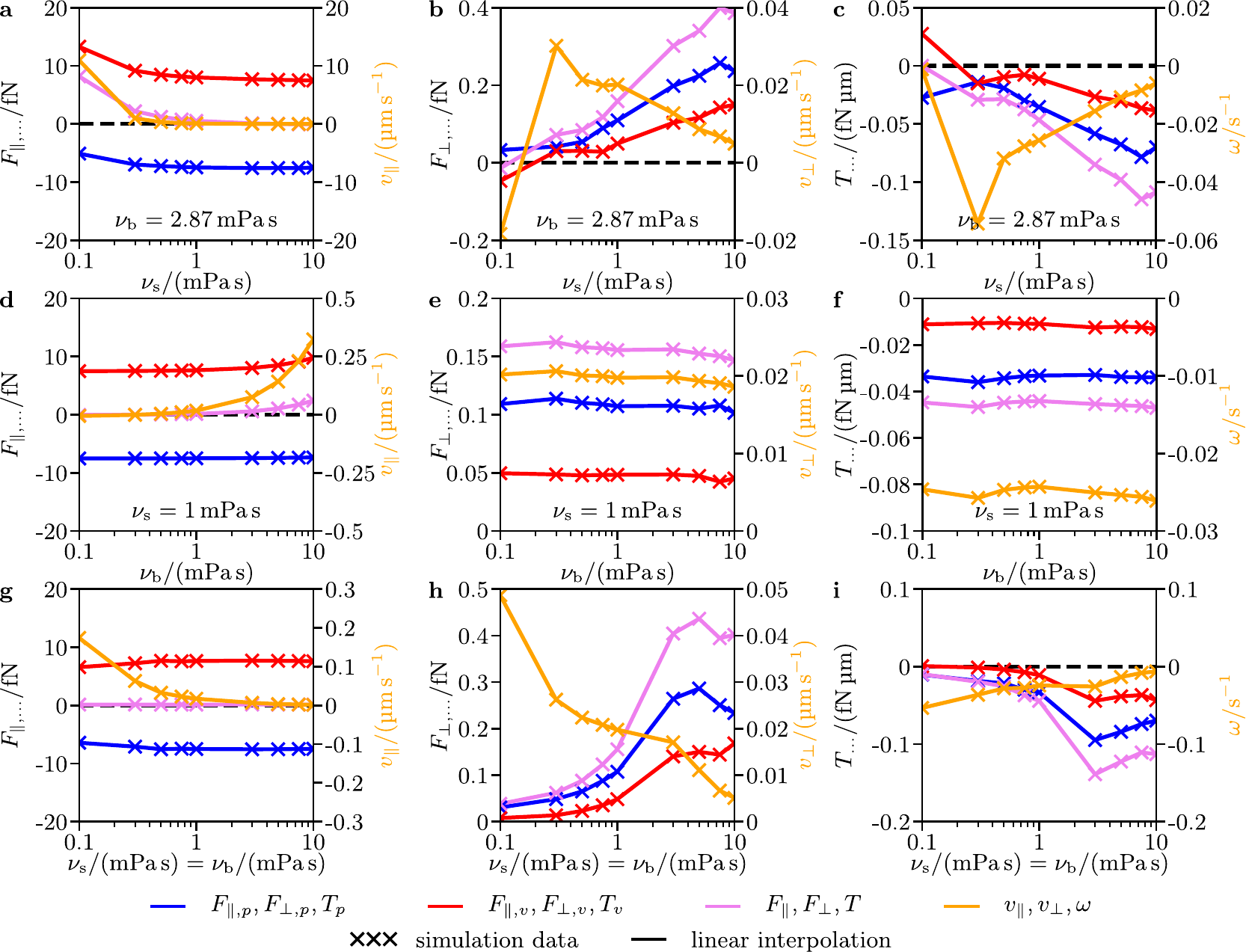}%
\caption{\label{fig:2}Simulation results for the pressure and viscous force contributions $F_{\parallel,p}$, $F_{\parallel,v}$, $F_{\perp,p}$, and $F_{\perp,v}$ and torque contributions $T_{p}$ and $T_{v}$ acting on the particle shown in Fig.\ \ref{fig:setup}, the parallel and perpendicular propulsion forces $F_\parallel=F_{\parallel,p}+F_{\parallel,v}$ and $F_\perp=F_{\perp,p}+F_{\perp,v}$, respectively, the propulsion torque $T=T_{p} + T_{v}$, the parallel and perpendicular propulsion velocities $v_\parallel$ and $v_\perp$, and the angular propulsion velocity $\omega$ for varying \textbf{a}-\textbf{c} shear viscosity $\nu_\mathrm{s}$, \textbf{d}-\textbf{f} bulk viscosity $\nu_\mathrm{b}$, and \textbf{g}-\textbf{i} shear and bulk viscosities $\nu_\mathrm{s}=\nu_\mathrm{b}$.}
\end{figure*}

The results of our simulations for the propulsion of the particle are shown in Fig.\ \ref{fig:2}.

\subsubsection{Variation of shear viscosity}
We start with varying the shear viscosity $\nu_\mathrm{s}\in[0.1,10]\,\SI{}{\milli\pascal\,\second}$ while keeping the bulk viscosity constant at $\nu_\mathrm{b}=\SI{2.87}{\milli\pascal\,\second}$ (see Fig.\ \ref{fig:2}\textbf{a}-\textbf{c}).

First, we consider the parallel components of the propulsion. 
The force components $F_{\parallel,p}$ and $F_{\parallel,v}$ as well as the parallel propulsion force $F_\parallel$ and the parallel propulsion velocity $v_\parallel$ decrease when $\nu_\mathrm{s}$ increases. 
$F_{\parallel,p}$ decreases from $F_{\parallel,p}=\SI{-5.10}{\femto\newton}$ to $F_{\parallel,p}=\SI{-7.50}{\femto\newton}$, $F_{\parallel,v}$ decreases from $F_{\parallel,v}=\SI{13.28}{\femto\newton}$ to $F_{\parallel,v}=\SI{7.48}{\femto\newton}$, 
$F_\parallel$ decreases from $F_\parallel=\SI{8.18}{\femto\newton}$ to $F_\parallel=\SI{-0.02}{\femto\newton}$, and $v_\parallel$ decreases from $v_\parallel=\SI{10.94}{\micro\metre\,\second^{-1}}$ to $v_\parallel=\SI{-0.0003}{\micro\metre\,\second^{-1}}$.
Remarkably, the decrease for the parallel propulsion involves a sign change, i.e., a thrust reversal of the particle.
The change of sign occurs between $\nu_\mathrm{s}=\SI{5}{\milli\pascal\,\second}$ and $\nu_\mathrm{s}=\SI{7.5}{\milli\pascal\,\second}$, i.e., for relatively large values of $\nu_\mathrm{s}$. 
According to our simulation data, the parallel propulsion force $F_\parallel$ scales roughly as $F_\parallel \sim 1/\nu_\mathrm{s}$, and the parallel propulsion velocity $v_\parallel$ scales roughly as $v_\parallel\sim 1/\nu_\mathrm{s}^2$. 
The faster decrease of $v_\parallel$ can be understood from the Stokes law \eqref{eq:velocity}, which implies $v_\parallel\sim F_\parallel/\nu_\mathrm{s}$.

This decrease of the propulsion for increasing shear viscosity $\nu_\mathrm{s}$ is consistent with the weakening of the flow field that we observed in Section \ref{sec:flowfields}. 
It is also in line with the findings of previous experiments \cite{GarciaGradillaEtAl2013,WangLMAHM2014,WuEtAl2014,EstebanEtAl2018,GaoLWWXH2019}, but the comparability is very limited due to the fact that in experiments the viscosities cannot be varied independently of other parameters of the system.
The observation of the reversal of the propulsion direction is consistent with the theory of Ref.\ \cite{CollisCS2017}.
This reference predicts a propulsion reversal at $\beta\in \mathcal{O}(1)$ with the 
acoustic Reynolds number $\beta = \pi \rho_0 \sigma^2 f/(2\nu_\mathrm{s})$.    
In our work, $\beta$ ranges in the interval $[0.0785,7.85]$ so that the observation of a sign change in our study is likely according to Ref.\ \cite{CollisCS2017}. 
Actually, the sign change occurs here in a range for the shear viscosity that corresponds to an acoustic Reynolds number between $\beta=0.2$ and $\beta=0.3$. 
 
Second, we consider the perpendicular components of the propulsion. 
Except for small nonmonotonicities, the force components $F_{\perp,p}$ and $F_{\perp,v}$ as well as the perpendicular propulsion force $F_\perp$ increase with $\nu_\mathrm{s}$.
$F_{\perp,p}$ increases from $F_{\perp,p}=\SI{0.033}{\femto\newton}$ to $F_{\perp,p}=\SI{0.24}{\femto\newton}$, $F_{\perp,v}$ increases from $F_{\perp,v}=\SI{-0.047}{\femto\newton}$ to $F_{\perp,v}=\SI{0.15}{\femto\newton}$, and 
$F_\perp$ increases from $F_\perp=\SI{-0.014}{\femto\newton}$ to $F_\perp=\SI{0.39}{\femto\newton}$.
In contrast, the perpendicular propulsion velocity $v_\perp$ first increases from $v_\perp=\SI{-0.018}{\micro\metre\,\second^{-1}}$ to $v_\perp=\SI{0.030}{\micro\metre\,\second^{-1}}$ at $\nu_\mathrm{s}=\SI{0.3}{\milli\pascal\,\second}$ and then decreases to $v_\perp=\SI{0.005}{\micro\metre\,\second^{-1}}$. 
This behavior of $v_\perp$ can be understood from the scaling $v_\perp\sim F_\perp/\nu_\mathrm{s}$ that follows from the Stokes law \eqref{eq:velocity}. 
As we can see, there is now a sign change of the perpendicular propulsion that occurs at low values of $\nu_\mathrm{s}$.

The observed sign change of the perpendicular propulsion can be explained by the balance of the acoustic radiation force and the acoustic streaming force that act on the particle in the perpendicular direction and occur also for highly symmetric particles including spheres \cite{Doinikov1994,Doinikov1997}. 
While the acoustic radiation force results from the scattering of the ultrasound wave at the particle, the acoustic streaming force corresponds to fluid streaming that is caused by the dissipation of energy of the ultrasound wave into the fluid. The acoustic streaming can occur even if there is no particle in the fluid \cite{Doinikov1994}.  
Both forces typically point in opposite directions \cite{Doinikov1997}, with the acoustic radiation force being parallel to the propagation direction of the ultrasound wave and the acoustic streaming force being antiparallel to the propagation direction for a spherical particle, and it is often difficult to predict which one dominates. 
For a spherical particle, the acoustic radiation force scales as $\sim \sqrt{\nu_\mathrm{s}}$ for small $\nu_\mathrm{s}$ and as $\sim \nu_\mathrm{s}$ for large $\nu_\mathrm{s}$ \cite{SettnesB2012}, and the acoustic streaming force scales as $\sim \nu_\mathrm{s}$ \cite{Doinikov1994,Bruus2012acoustofluidics}.
This shows that the sign of the total force acting on the particle in the perpendicular direction, i.e., parallel to the propagation direction of the ultrasound, can change when the value of the shear viscosity $\nu_\mathrm{s}$ is varied. 
(Whether the sign actually changes depends on the shape of the particle.) 
 
Third, we consider the angular components of the propulsion.
The torque components $T_p$ and $T_v$, as well as the propulsion torque $T$, have a clear downward trend for increasing $\nu_\mathrm{s}$, however, with significant nonmonotonicities.
$T_p$ decreases from $T_p=\SI{-0.03}{\femto\newton\,\micro\metre}$ to $T_p=\SI{-0.07}{\femto\newton\,\micro\metre}$, $T_v$ decreases from $T_v=\SI{0.03}{\femto\newton\,\micro\metre}$ to $T_v=\SI{-0.04}{\femto\newton\,\micro\metre}$, and $T$ decreases from zero to $T=\SI{-0.11}{\femto\newton\,\micro\metre}$.
In contrast, the angular propulsion velocity $\omega$ shows no clear downward or upward trend.
It decreases from $\omega=\SI{-0.001}{\second^{-1}}$ to $\omega=\SI{-0.054}{\second^{-1}}$ at $\nu_\mathrm{s}=\SI{0.3}{\milli\pascal\,\second}$ and then increases to $\omega=\SI{-0.006}{\second^{-1}}$. 
Similar to the behavior of $v_\perp$, the behavior of $\omega$ can be understood from the scaling $\omega\sim T/\nu_\mathrm{s}$ that follows from the Stokes law \eqref{eq:velocity}.
We thus see that different from translational propulsion, angular propulsion involves no change of sign.

\subsubsection{Variation of bulk viscosity}
Next, we vary the bulk viscosity $\nu_\mathrm{b}\in[0.1,10]\,\SI{}{\milli\pascal\,\second}$ while keeping the shear viscosity constant at $\nu_\mathrm{s}=\SI{1}{\milli\pascal\,\second}$ (see Fig.\ \ref{fig:2}\textbf{d}-\textbf{f}).

First, we consider the parallel components of the propulsion. 
The force component $F_{\parallel,p}$ is rather independent of $\nu_\mathrm{b}$. It only increases from $F_{\parallel,p}=\SI{-7.53}{\femto\newton}$ to $F_{\parallel,p}=\SI{-7.37}{\femto\newton}$ when $\nu_\mathrm{b}$ is increased. 
For the force component $F_{\parallel,v}$, on the other hand, we see that its value remains rather constant for small values of $\nu_\mathrm{b}$ but significantly increases for $\nu_\mathrm{b} > \SI{1}{\milli\pascal\,\second}$.
Its total increase reaches from $F_{\parallel,v}=\SI{7.49}{\femto\newton}$ to $F_{\parallel,v}=\SI{9.79}{\femto\newton}$.
For the parallel propulsion force $F_\parallel$ and the parallel propulsion velocity $v_\parallel$, we observe a similar behavior as for $F_{\parallel,v}$, but the increase for $\nu_\mathrm{b} > \SI{1}{\milli\pascal\,\second}$ is more pronounced for $F_\parallel$ and even more for $v_\parallel$. 
$F_\parallel$ increases from $F_\parallel=\SI{-0.04}{\femto\newton}$ to $F_\parallel=\SI{2.42}{\femto\newton}$ and $v_\parallel$ increases from $v_\parallel=\SI{-0.006}{\micro\metre\,\second^{-1}}$ to $v_\parallel=\SI{0.32}{\micro\metre\,\second^{-1}}$. 
Similar to a variation of the shear viscosity $\nu_\mathrm{s}$, we now see that the parallel propulsion changes sign when the bulk viscosity $\nu_\mathrm{b}$ is varied. 
The sign change occurs between $\nu_\mathrm{b}=\SI{0.3}{\milli\pascal\,\second}$ and $\nu_\mathrm{b}=\SI{0.5}{\milli\pascal\,\second}$. 
However, the overall dependence of the parallel propulsion on $\nu_\mathrm{b}$ is rather different from its dependence on $\nu_\mathrm{s}$. 
While the parallel propulsion increases with $\nu_\mathrm{b}$, it decreases for increasing $\nu_\mathrm{s}$. 
Here, we cannot compare our results to experimental, numerical, or analytical studies from the literature, since it seems that no previous findings on the dependence of the parallel acoustic propulsion on the bulk viscosity $\nu_\mathrm{b}$ have been published. 
 
Now, we consider the perpendicular and angular components of the propulsion. 
Interestingly, all these components are rather independent of the bulk viscosity $\nu_\mathrm{b}$.
Their values are roughly constant at $F_{\perp,p}=\SI{0.11}{\femto\newton}$, $F_{\perp,v}=\SI{0.05}{\femto\newton}$, $F_\perp=\SI{0.16}{\femto\newton}$, and $v_\perp=\SI{0.020}{\micro\metre\,\second^{-1}}$ 
as well as at $T_p=\SI{-0.034}{\femto\newton\,\micro\metre}$, $T_v=\SI{-0.012}{\femto\newton\,\micro\metre}$, $T=\SI{-0.046}{\femto\newton\,\micro\metre}$, and $\omega=\SI{-0.025}{\second^{-1}}$.   
The fact that the perpendicular propulsion is rather independent of $\nu_\mathrm{b}$ is in line with theoretical treatments of the force that ultrasound exerts on a sphere in the direction of propagation of the ultrasound wave. 
In previous theoretical approaches, the dependence of this force on the bulk viscosity $\nu_\mathrm{b}$ has been found to be negligibly small \cite{Doinikov1994} or it has directly been ignored \cite{Doinikov1997}. 
 
We thus see that the bulk viscosity $\nu_\mathrm{b}$ has a much smaller influence on the propulsion of the particle than the shear viscosity $\nu_\mathrm{s}$. This agrees with the weak dependence of the particle's flow field on $\nu_\mathrm{b}$ that we observed in Section \ref{sec:flowfields}.

\subsubsection{Variation of shear and bulk viscosities}
Finally, we vary both viscosities jointly as $\nu_\mathrm{s}=\nu_\mathrm{b}\in[0.1,10]\,\SI{}{\milli\pascal\,\second}$ (see Fig.\ \ref{fig:2}\textbf{g}-\textbf{i}). 

The parallel force components $F_{\parallel,p}$ and $F_{\parallel,v}$ and the parallel propulsion force $F_\parallel$ remain rather constant at $F_{\parallel,p}=\SI{-7.35}{\femto\newton}$, $F_{\parallel,v}=\SI{7.48}{\femto\newton}$, and $F_\parallel=\SI{0.13}{\femto\newton}$ when the viscosities are increased. 
As a consequence of the constant parallel propulsion force for increasing viscosities, the parallel propulsion velocity $v_\parallel$ decreases from $v_\parallel=\SI{0.17}{\micro\metre\,\second^{-1}}$ to $v_\parallel=\SI{0.001}{\micro\metre\,\second^{-1}}$. 
%
The perpendicular force components $F_{\perp,p}$ and $F_{\perp,v}$ and the perpendicular propulsion force $F_\perp$ increase with the viscosities from $F_{\perp,p}=\SI{0.030}{\femto\newton}$, $F_{\perp,v} = \SI{0.008}{\femto\newton}$, and $F_\perp=\SI{0.038}{\femto\newton}$ until $F_{\perp,p}=\SI{0.29}{\femto\newton}$, $F_{\perp,v}=\SI{0.15}{\femto\newton}$, and $F_\perp=\SI{0.44}{\femto\newton}$ at about $\nu_\mathrm{s}=\nu_\mathrm{b}=\SI{5}{\milli\pascal\,\second}$ and afterward slightly decrease or increase to $F_{\perp,p}=\SI{0.23}{\femto\newton}$, $F_{\perp,v}=\SI{0.17}{\femto\newton}$, and $F_\perp=\SI{0.40}{\femto\newton}$. 
In contrast, the perpendicular propulsion velocity $v_\perp$ decreases from $v_\perp=\SI{0.049}{\micro\metre\,\second^{-1}}$ to $v_\perp=\SI{0.005}{\micro\metre\,\second^{-1}}$, since the increase of the viscosities dominates the increase of the propulsion force.
%
The torque components $T_{p}$ and $T_{v}$ and the propulsion torque $T$ behave qualitatively similar when the viscosities are increased. 
They start with a value close to zero $T_{p}=\SI{-0.010}{\femto\newton\,\micro\metre}$, $T_{v}=\SI{0.001}{\femto\newton\,\micro\metre}$, and $T=\SI{-0.01}{\femto\newton\,\micro\metre}$, decrease until $T_{p}=\SI{-0.095}{\femto\newton\,\micro\metre}$, $T_{v}=\SI{-0.044}{\femto\newton\,\micro\metre}$, and $T=\SI{-0.139}{\femto\newton\,\micro\metre}$ at about $\nu_\mathrm{s}=\nu_\mathrm{b}=\SI{3}{\milli\pascal\,\second}$, and afterward increase to $T_{p}=\SI{-0.070}{\femto\newton\,\micro\metre}$, $T_{v}=\SI{-0.043}{\femto\newton\,\micro\metre}$, and $T=\SI{-0.113}{\femto\newton\,\micro\metre}$. 
For the angular propulsion velocity $\omega$, we find a rather linear increase with the viscosities from $\omega=\SI{-0.05}{\second^{-1}}$ to $\omega=\SI{-0.006}{\second^{-1}}$. 
 
Overall, the behavior of the propulsion for a joint variation of both viscosities shows some parallels to the behavior for a variation of only the shear viscosity $\nu_\mathrm{s}$, but there are also significant differences. This is interesting, as we observed only a very weak dependence of the propulsion on the bulk viscosity $\nu_\mathrm{b}$.

\subsection{Viscosity-dependent motion}
Finally, we characterize the particle motion by comparing its Brownian rotation with its translational and rotational propulsion and study how the type of motion depends on the shear viscosity $\nu_\mathrm{s}$ and bulk viscosity $\nu_\mathrm{b}$ as well as on the acoustic energy density $E$.
Using the classification for the type of motion described in Section \ref{sec:characterization}, we can distinguish \ZT{random motion} ($E < \min\{E_\mathrm{dir},E_\mathrm{gui}\}$, Brownian rotation dominates translational and rotational propulsion), \ZT{directional motion} ($E > \min\{E_\mathrm{dir},E_\mathrm{gui}\}$, translational or rotational propulsion dominates Brownian rotation), \ZT{random orientation} ($E < E_\mathrm{gui}$, Brownian rotation dominates rotational propulsion), and \ZT{guided motion} ($E > E_\mathrm{gui}$, rotational propulsion dominates Brownian rotation), where the energy density thresholds $E_\mathrm{dir}$ and $E_\mathrm{gui}$ are defined by Eqs.\ \eqref{eq:E_dir_lim} and \eqref{eq:E_Br_lim}.  
These thresholds depend on the particle's rotational diffusion coefficient $D_\mathrm{R}=(k_\mathrm{B} T_0 / \nu_\mathrm{s}) (\boldsymbol{\mathrm{H}}^{-1})_{66}$, which in turn depends on the shear viscosity $\nu_\mathrm{s}$, as well as on the parallel propulsion velocity $v_\parallel$ and the angular propulsion velocity $\omega$.  
To determine the values of $v_\parallel$ and $\omega$ that occur in Eqs.\ \eqref{eq:E_dir_lim} and \eqref{eq:E_Br_lim} for other acoustic energy densities $E$ than $E_\mathrm{R}$, which corresponds to our simulations, we rescaled our results for $v_\parallel$ and $\omega$ whilst taking into account that these parameters are approximately proportional to $E$ \cite{VossW2022acoustic}. 

\begin{figure*}[htb]
\centering
\includegraphics[width=\linewidth]{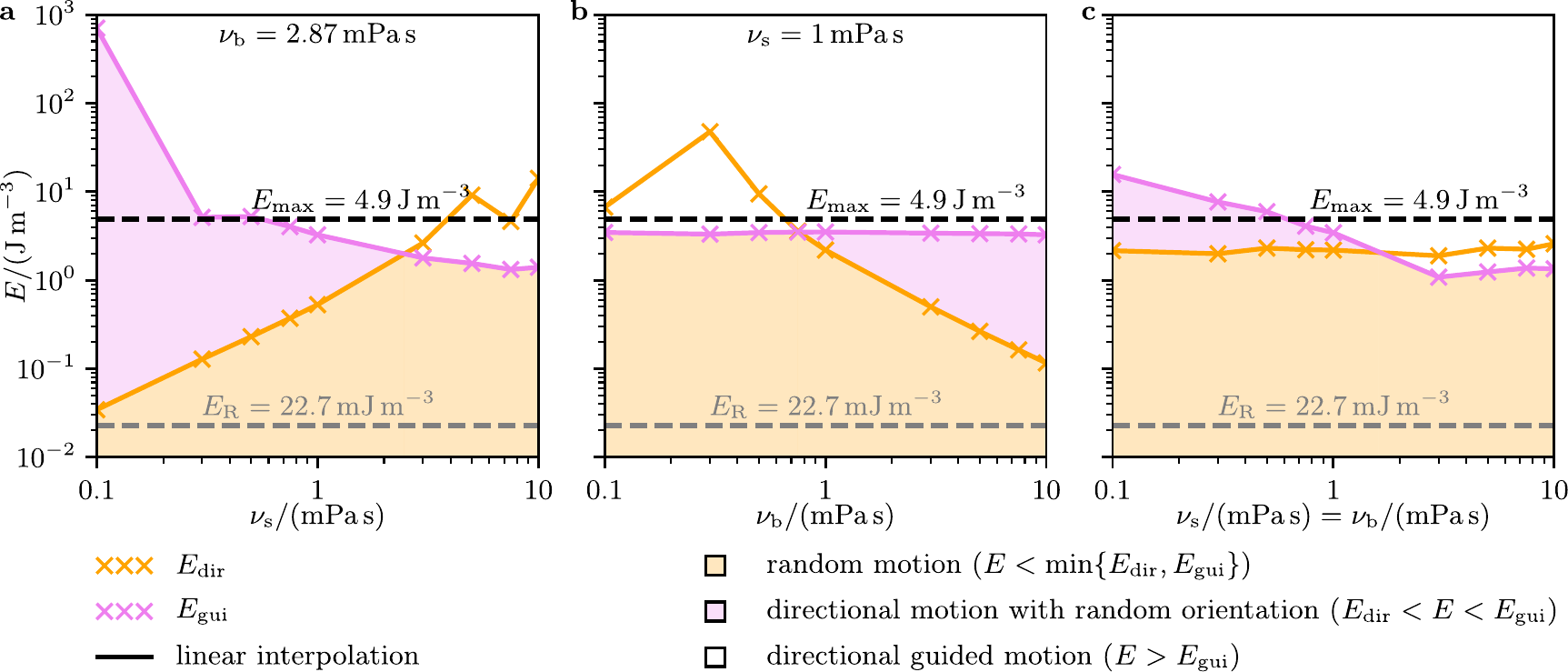}%
\caption{\label{fig:energy}Characterization of the qualitative particle motion according to Section \ref{sec:characterization} for varying acoustic energy density $E$ and \textbf{a} shear viscosity $\nu_\mathrm{s}$, \textbf{b} bulk viscosity $\nu_\mathrm{b}$, and \textbf{c} shear and bulk viscosities $\nu_\mathrm{s}=\nu_\mathrm{b}$. The acoustic energy density $E_\mathrm{R}$ used in our simulations and the maximal acoustic energy density $E_\mathrm{max}$ that is suitable for diagnostic applications in the human body \cite{BarnettEtAl2000} are indicated.}
\end{figure*}

Figure \ref{fig:energy} shows which types of motion can be expected depending on the values of $\nu_\mathrm{s}$, $\nu_\mathrm{b}$, and $E$. 
We see that for very low values of the acoustic energy density $E$, the particle performs a random motion, independent of the values of the shear viscosity $\nu_\mathrm{s}$ and bulk viscosity $\nu_\mathrm{b}$. 
For moderate values of $E$, we see that random motion and directional motion are possible.
Directional motion is preferred for small $\nu_\mathrm{s}$ and large $\nu_\mathrm{b}$. 
The threshold value $E=\min\{E_\mathrm{dir},E_\mathrm{gui}\}$, where random motion is replaced by directional motion, increases with $\nu_\mathrm{s}$ when $\nu_\mathrm{b}$ is kept constant, decreases with $\nu_\mathrm{b}$ (except for very low values of $\nu_\mathrm{b}$) when $\nu_\mathrm{s}$ is kept constant, and is nearly independent of $\nu_\mathrm{s}=\nu_\mathrm{b}$ when both viscosities are varied simultaneously.
In the region for directional motion, directional motion with random orientation and guided motion can be distinguished. 
While the former type of motion can occur for moderate values of $E$, the latter type of motion is found when $E$ is large. 
The threshold value $E=E_\mathrm{gui}$, where directional motion with a random orientation is replaced by guided motion, decreases with $\nu_\mathrm{s}$, irrespective of whether $\nu_\mathrm{b}$ is kept constant or not, and is independent of $\nu_\mathrm{b}$ when $\nu_\mathrm{s}$ is kept constant. 
Therefore, directional motion with a random orientation does not occur at all for large $\nu_\mathrm{s}$, when $\nu_\mathrm{b}$ is kept constant, for small $\nu_\mathrm{b}$, when $\nu_\mathrm{s}$ is kept constant, and for large $\nu_\mathrm{s}=\nu_\mathrm{b}$, when both viscosities are varied simultaneously. Instead, for increasing $E$ random motion is directly replaced by guided motion in these cases.
When the particle is propelled by ultrasound with the maximal acoustic energy density $E_\mathrm{max}$ that is suitable for diagnostic applications in the human body \cite{BarnettEtAl2000}, its motion is always directional.

\section{\label{conclusions}Conclusions}
We have investigated how the acoustic propulsion of a cone-shaped particle in a planar traveling ultrasound wave depends on the shear and bulk viscosities of the fluid surrounding the particle. 
Our study addressed the flow field generated by the particle, the resulting propulsion force and torque, the corresponding translational and angular propulsion velocities, and the type of motion that corresponds to the propulsion. In the latter case, we took also the effect of a variation of the acoustic energy density of the ultrasound field into account. 
As a main result, we observed that the propulsion becomes weaker and can even change sign for increasing shear viscosity and that the propulsion is rather independent of the bulk viscosity. 

The detailed analysis of the effect of the fluid's viscosities on the particle's propulsion is important progress in the investigation of acoustically propelled nano- and microparticles since insights into how the viscosity affects the acoustic propulsion have been very limited for the shear viscosity and even unavailable for the bulk viscosity up to now \cite{GarciaGradillaEtAl2013,WuEtAl2014,EstebanEtAl2018,GaoLWWXH2019,WangLMAHM2014,CollisCS2017}.
Reasons for this are that it is extremely difficult or even impossible to vary one of these viscosities in experiments without changing other parameters of the system and that previous theory-based approaches have neglected the bulk viscosity \cite{Zhou2018,SabrinaTABdlCMB2018,WangGWSGXH2018,TangEtAl2019,RenEtAl2019,CollisCS2017,NadalL2014}. 
While previous studies usually considered water as the fluid surrounding the particles \cite{WangCHM2012,AhmedEtAl2013,BalkEtAl2014,AhmedGFM2014,SotoWGGGLKACW2016,AhmedWBGHM2016,AhmedBJPDN2016,ZhouZWW2017,SabrinaTABdlCMB2018,WangGWSGXH2018,RenEtAl2019,VossW2020,ValdezLOESSWG2020,AghakhaniYWS2020,LiuR2020,DumyJMBGMHA2020,VossW2021,VossW2022orientation,VossW2022acoustic,McneillNBM2020,MohantyEtAl2021,McneillSWOLNM2021}, future applications of acoustically propelled nano- and microparticles will involve other fluids \cite{GarciaGradillaEtAl2013,WuEtAl2014,WangLMAHM2014,EstebanEtAl2018,GaoLWWXH2019,WangGZLH2020,WuEtAl2015a,EstebanFernandezdeAvilaMSLRCVMGZW2015,EstebanEtAl2016,EstebanFernandezEtAl2017,UygunEtAl2017,HansenEtAl2018,QualliotineEtAl2019}, which will have other viscosities. 
Important examples are biofluids since medicine is the most prominent field of application of acoustically propelled particles that has been envisaged \cite{LiEFdAGZW2017,PengTW2017,SotoC2018,WangGZLH2020,WangZ2021,LuoFWG2018,ErkocYCYAS2019}. For future research, it would be interesting to study the viscosity-dependence of particles with other shapes and to consider also acoustically propelled particles in non-Newtonian fluids.

\section*{Data availability}
The raw data corresponding to the figures shown in this article are available as Supplementary Material \cite{SI}.

\section*{Conflicts of interest}
There are no conflicts of interest to declare.

\begin{acknowledgments}
We thank Patrick Kurzeja for helpful discussions. 
R.W.\ is funded by the Deutsche Forschungsgemeinschaft (DFG, German Research Foundation) -- WI 4170/3-1. 
The simulations for this work were performed on the computer cluster PALMA II of the University of M\"unster. 
\end{acknowledgments}

\nocite{apsrev41Control}
\bibliographystyle{apsrev4-1}
\bibliography{control,refs}
	
\end{document}